\documentclass[twocolumn]{aastex631}
\shorttitle{A High Density of Bright Galaxies at $z\approx10$ in the A2744 region}
\shortauthors{Castellano et al.}
\usepackage{hyperref}  
\usepackage{amsmath} 
\usepackage{rotating,tabularx}
\hypersetup{colorlinks=true,linkcolor=[rgb]{1.,0.2,0.2},citecolor=[rgb]{0.1,0.4,1.},filecolor=[rgb]{0.7,0.2,0.2},urlcolor=[rgb]{0.7,0.2,0.2}}

\usepackage{color}
\usepackage{makecell}
\usepackage{graphicx}
\usepackage{multirow}
\usepackage{txfonts}
\definecolor{blue}{rgb}{0., 0., 1}

\defcitealias{Bergamini2022}{B23}

\begin{document}

\title{Early Results from GLASS-JWST. XIX: A High Density of Bright Galaxies at $z\approx10$ in the Abell 2744 Region
}

\correspondingauthor{Marco Castellano}
\email{marco.castellano@inaf.it}

\author[0000-0001-9875-8263]{Marco Castellano}
\affiliation{INAF - Osservatorio Astronomico di Roma, via di Frascati 33, 00078 Monte Porzio Catone, Italy}

 \author[0000-0003-3820-2823]{Adriano Fontana}
 \affiliation{INAF - Osservatorio Astronomico di Roma, via di Frascati 33, 00078 Monte Porzio Catone, Italy}

 \author[0000-0002-8460-0390]{Tommaso Treu}
 \affiliation{Department of Physics and Astronomy, University of California, Los Angeles, 430 Portola Plaza, Los Angeles, CA 90095, USA}

\author[0000-0001-6870-8900]{Emiliano Merlin}
\affiliation{INAF - Osservatorio Astronomico di Roma, via di Frascati 33, 00078 Monte Porzio Catone, Italy}

 \author[0000-0002-9334-8705]{Paola Santini}
 \affiliation{INAF - Osservatorio Astronomico di Roma, via di Frascati 33, 00078 Monte Porzio Catone, Italy}

\author[0000-0003-1383-9414]{Pietro Bergamini}
\affiliation{Dipartimento di Fisica, Università degli Studi di Milano, via Celoria 16, I-20133 Milano, Italy}
\affiliation{INAF - OAS, Osservatorio di Astrofisica e Scienza dello Spazio di Bologna, via Gobetti 93/3, I-40129 Bologna, Italy}

\author[0000-0002-5926-7143]{Claudio Grillo}
 \affiliation{Dipartimento di Fisica, Università degli Studi di Milano, via Celoria 16, I-20133 Milano, Italy}
 \affiliation{INAF—IASF Milano, via A. Corti 12, I-20133 Milano, Italy}
 
\author[0000-0002-6813-0632]{Piero Rosati}
\affiliation{Dipartimento di Fisica e Scienze della Terra, Università degli Studi di Ferrara, Via Saragat 1, I-44122 Ferrara, Italy}
\affiliation{INAF - OAS, Osservatorio di Astrofisica e Scienza dello Spazio di Bologna, via Gobetti 93/3, I-40129 Bologna, Italy}

\author[0000-0002-5926-7143]{Ana Acebron}
 \affiliation{Dipartimento di Fisica, Università degli Studi di Milano, via Celoria 16, I-20133 Milano, Italy}
 \affiliation{INAF—IASF Milano, via A. Corti 12, I-20133 Milano, Italy}

\author[0000-0003-4570-3159]{Nicha Leethochawalit}
\affiliation{National Astronomical Research Institute of Thailand (NARIT), Mae Rim, Chiang Mai, 50180, Thailand}

\author[0000-0002-7409-8114]{Diego Paris}
 \affiliation{INAF - Osservatorio Astronomico di Roma, via di Frascati 33, 00078 Monte Porzio Catone, Italy}

 \author{Andrea Bonchi}
 \affiliation{Space Science Data Center, Italian Space Agency, via del Politecnico, 00133, Roma, Italy}

 \author{Davide Belfiori}
 \affiliation{INAF - Osservatorio Astronomico di Roma, via di Frascati 33, 00078 Monte Porzio Catone, Italy}

 \author[0000-0003-2536-1614]{Antonello Calabr\`o}
 \affiliation{INAF - Osservatorio Astronomico di Roma, via di Frascati 33, 00078 Monte Porzio Catone, Italy}

\author{Matteo Correnti}
 \affiliation{Space Science Data Center, Italian Space Agency, via del Politecnico, 00133, Roma, Italy}
 
 \author[0000-0001-6342-9662]{Mario Nonino}
 \affiliation{INAF - Osservatorio Astronomico di Trieste, Via Tiepolo 11, I-34131 Trieste, Italy}

 \author[0000-0003-4067-9196]{Gianluca Polenta}
 \affiliation{Space Science Data Center, Italian Space Agency, via del Politecnico, 00133, Roma, Italy}

\author[0000-0001-9391-305X]{Michele Trenti}
\affiliation{School of Physics, University of Melbourne, Parkville 3010, VIC, Australia}
\affiliation{ARC Centre of Excellence for All Sky Astrophysics in 3 Dimensions (ASTRO 3D), Australia}

\author[0000-0003-4109-304X]{Kristan Boyett}
\affiliation{School of Physics, University of Melbourne, Parkville 3010, VIC, Australia}
\affiliation{ARC Centre of Excellence for All Sky Astrophysics in 3 Dimensions (ASTRO 3D), Australia}

\author[0000-0003-2680-005X]{G. Brammer}
\affiliation{Cosmic Dawn Center (DAWN), Denmark}
\affiliation{Niels Bohr Institute, University of Copenhagen, Jagtvej 128, DK-2200 Copenhagen N, Denmark}

\author[0000-0002-5807-4411]{Tom Broadhurst}
\affiliation{Department of Theoretical Physics, University of the Basque Country UPV-EHU, 48040 Bilbao, Spain}
\affiliation{Donostia International Physics Center (DIPC), 20018 Donostia, The Basque Country}
\affiliation{IKERBASQUE, Basque Foundation for Science, Alameda Urquijo, 36-5 48008 Bilbao, Spain}

\author[0000-0001-6052-3274]{Gabriel B. Caminha}
\affiliation{Technische Universität München, Physik-Department, James-Franck Str. 1, 85748 Garching, Germany}
\affiliation{Max-Planck-Institut f\"ur Astrophysik, Karl-Schwarzschild-Str. 1, D-85748 Garching, Germany}

\author[0000-0003-1060-0723]{Wenlei Chen}
\affiliation{School of Physics and Astronomy, University of Minnesota, 116 Church Street SE, Minneapolis, MN 55455, USA}

\author[0000-0003-3460-0103]{Alexei V. Filippenko}
\affiliation{Department of Astronomy, University of California, Berkeley, CA 94720-3411, USA}

\author[0000-0001-6793-2572]{Flaminia Fortuni}
 \affiliation{INAF - Osservatorio Astronomico di Roma, via di Frascati 33, 00078 Monte Porzio Catone, Italy}

\author[0000-0002-3254-9044]{Karl Glazebrook}
\affiliation{Centre for Astrophysics and Supercomputing, Swinburne University of Technology, PO Box 218, Hawthorn, VIC 3122, Australia}

\author[0000-0002-9572-7813]{Sara Mascia}
\affiliation{INAF - Osservatorio Astronomico di Roma, via di Frascati 33, 00078 Monte Porzio Catone, Italy}

\author[0000-0002-3407-1785]{Charlotte A. Mason}
\affiliation{Cosmic Dawn Center (DAWN)}
\affiliation{Niels Bohr Institute, University of Copenhagen, Jagtvej 128, 2200 København N, Denmark}

\author[0000-0002-4096-2680]{Nicola Menci}
 \affiliation{INAF - Osservatorio Astronomico di Roma, via di Frascati 33, 00078 Monte Porzio Catone, Italy}

\author[0000-0003-1225-7084]{Massimo Meneghetti}
\affiliation{INAF -- OAS, Osservatorio di Astrofisica e Scienza dello Spazio di Bologna, via Gobetti 93/3, I-40129 Bologna, Italy}
\affiliation{INFN-Sezione di Bologna, Viale Berti Pichat 6/2, I-40127 Bologna, Italy}

\author[0000-0001-9261-7849]{Amata Mercurio}
\affiliation{Dipartimento di Fisica “E.R. Caianiello”, Universit\`a Degli Studi di Salerno, Via Giovanni Paolo II, I–84084 Fisciano (SA), Italy}
\affiliation{INAF - Osservatorio Astronomico di Capodimonte, Via Moiariello 16, 80131 Napoli, Italy}

\author[0000-0002-8632-6049]{Benjamin Metha}
\affiliation{School of Physics, University of Melbourne, Parkville 3010, VIC, Australia}
\affiliation{ARC Centre of Excellence for All Sky Astrophysics in 3 Dimensions (ASTRO 3D), Australia}
\affiliation{Department of Physics and Astronomy, University of California, Los Angeles, 430 Portola Plaza, Los Angeles, CA 90095, USA}

\author[0000-0002-8512-1404]{Takahiro Morishita}
\affiliation{IPAC, California Institute of Technology, MC 314-6, 1200 E. California Boulevard, Pasadena, CA 91125, USA}

\author[0000-0003-2804-0648 ]{Themiya Nanayakkara}
\affiliation{Centre for Astrophysics and Supercomputing, Swinburne University of Technology, PO Box 218, Hawthorn, VIC 3122, Australia}

\author[0000-0001-8940-6768 ]{Laura Pentericci}
\affiliation{INAF - Osservatorio Astronomico di Roma, via di Frascati 33, 00078 Monte Porzio Catone, Italy}

\author[0000-0002-4140-1367]{Guido Roberts-Borsani}
\affiliation{Department of Physics and Astronomy, University of California, Los Angeles, 430 Portola Plaza, Los Angeles, CA 90095, USA}

\author[0000-0002-4430-8846]{Namrata Roy}
\affiliation{Center for Astrophysical Sciences, Department of Physics and Astronomy, The Johns Hopkins University, Baltimore, MD 21218, USA}

\author[0000-0002-5057-135X]{Eros Vanzella}
\affiliation{INAF -- OAS, Osservatorio di Astrofisica e Scienza dello Spazio di Bologna, via Gobetti 93/3, I-40129 Bologna, Italy}

\author[0000-0003-0980-1499]{Benedetta Vulcani}
\affiliation{INAF Osservatorio Astronomico di Padova, vicolo dell'Osservatorio 5, 35122 Padova, Italy}

\author[0000-0002-8434-880X]{Lilan Yang}
\affiliation{Kavli Institute for the Physics and Mathematics of the Universe, The University of Tokyo, Kashiwa, Japan 277-8583}

\author[0000-0002-9373-3865]{Xin Wang}
\affil{School of Astronomy and Space Science, University of Chinese Academy of Sciences (UCAS), Beijing 100049, China}
\affil{National Astronomical Observatories, Chinese Academy of Sciences, Beijing 100101, China}
\affil{Institute for Frontiers in Astronomy and Astrophysics, Beijing Normal University,  Beijing 102206, China}

% \author{xxx}
% \affiliation{xx, USA}

%\collaboration{6}{(AAS Journals Data Editors)}
%
%\author{Butler Burton}
%\affiliation{Leiden University}
%\affiliation{AAS Journals Associate Editor-in-Chief}
%
%\author{Amy Hendrickson}
%\altaffiliation{AASTeX v6+ programmer}
%\affiliation{TeXnology Inc.}
%
%\author{Julie Steffen}
%\affiliation{AAS Director of Publishing}
%\affiliation{American Astronomical Society \\
%1667 K Street NW, Suite 800 \\
%Washington, DC 20006, USA}

%\author{Magaret Donnelly}
%\affiliation{IOP Publishing, Washington, DC 20005}

%% Note that the \and command from previous versions of AASTeX is now
%% depreciated in this version as it is no longer necessary. AASTeX 
%% automatically takes care of all commas and "and"s between authors names.

%% AASTeX 6.31 has the new \collaboration and \nocollaboration commands to
%% provide the collaboration status of a group of authors. These commands 
%% can be used either before or after the list of corresponding authors. The
%% argument for \collaboration is the collaboration identifier. Authors are
%% encouraged to surround collaboration identifiers with ()s. The 
%% \nocollaboration command takes no argument and exists to indicate that
%% the nearby authors are not part of surrounding collaborations.

%% Mark off the abstract in the ``abstract'' environment. 

\begin{abstract}
We report the detection of a high density of redshift $z\approx 10$ galaxies behind the foreground cluster Abell 2744, selected from imaging data obtained recently with NIRCam onboard {\it JWST} by three programs --- GLASS-JWST, UNCOVER, and DDT\#2756. To ensure robust estimates of the lensing magnification $\mu$, we use an improved version of our model that exploits the first epoch of NIRCam images and newly obtained MUSE spectra, and avoids regions with $\mu>5$ where the uncertainty may be higher. We detect seven bright $z\approx 10$ galaxies with demagnified rest-frame $-22 \lesssim M_{\rm UV}\lesssim -19$ mag, over an area of $\sim37$ sq. arcmin. Taking into account photometric incompleteness and the effects of lensing on luminosity and cosmological volume, we find that the density of $z\approx 10$ galaxies in the field is about $10\times$ ($3\times$) larger than the average at $M_{UV}\approx -21~ (-20)$ mag reported so far. The density is even higher when considering only the GLASS-JWST data, which are the deepest and the least affected by magnification and incompleteness. The GLASS-JWST field contains 5 out of 7 galaxies, distributed along an apparent filamentary structure of 2 Mpc in projected length, and includes a close pair of candidates with $M_{\rm UV}< -20$ mag having a projected separation of only 16 kpc. These findings suggest the presence of a $z\approx 10$ overdensity in the field. In addition to providing excellent targets for efficient spectroscopic follow-up observations, our study confirms the high density of bright galaxies observed in early {\it JWST} observations, but calls for multiple surveys along independent lines of sight to achieve an unbiased estimate of their average density and a first estimate of their clustering. 
\end{abstract}

%% Keywords should appear after the \end{abstract} command. 
%% The AAS Journals now uses Unified Astronomy Thesaurus concepts:
%% https://astrothesaurus.org
%% You will be asked to selected these concepts during the submission process
%% but this old "keyword" functionality is maintained in case authors want
%% to include these concepts in their preprints.

\keywords{Lyman-break galaxies --- Reionization --- Surveys}

%% From the front matter, we move on to the body of the paper.
%% Sections are demarcated by \section and \subsection, respectively.
%% Observe the use of the LaTeX \label
%% command after the \subsection to give a symbolic KEY to the
%% subsection for cross-referencing in a \ref command.
%% You can use LaTeX's \ref and \label commands to keep track of
%% cross-references to sections, equations, tables, and figures.
%% That way, if you change the order of any elements, LaTeX will
%% automatically renumber them.
%%
%% We recommend that authors also use the natbib \citep
%% and \citet commands to identify citations.  The citations are
%% tied to the reference list via symbolic KEYs. The KEY corresponds
%% to the KEY in the \bibitem in the reference list below. 

\section{Introduction}\label{sec:intro}

In just a few months, {\it JWST} has started to transform our understanding of the epoch of ``cosmic dawn," when the first sources of light likely started reionizing the intergalactic medium \citep{Dayal2018,Robertson2022a}. Previous {\it Hubble Space Telescope (HST)} and ground-based surveys enabled a census of galaxies  at redshift $z\approx6$--8, about 1~Gyr after the Big Bang \citep[e.g.,][]{Finkelstein2015,Castellano2016c,Bouwens2021}, and the first investigations of galaxies at $z \approx 9$--11 \citep[][]{Ellis2013,Bouwens2016b,McLeod2016,Ishigaki2018,Oesch2018,Morishita2018,Stefanon2019,Bowler2020,Roberts-Borsani2022,Leethochawalit2022a,Bagley2022}. The public datasets gathered through the {\it JWST} Early Release Observations \citep{Pontoppidan2022} and Early Release Science Programs have enabled the detection of tens of candidate sources at $z>
9$, pushing the cosmic frontier beyond the limits of HST-WFC3 to the first 200--300~Myr after the Big Bang \citep[e.g.,][]{Castellano2022b,Donnan2022,Finkelstein2022b,morishita22a,Naidu2022b,Yan2022,RobertsBorsani2022d,Robertson2022b,Bunker2023}. 

The results have been surprising: multiple independent analyses have shown that the number density of bright galaxies at $z>9$ is larger than predicted by theoretical models or on the basis of the extrapolation from lower-redshift estimates \citep[e.g.,][]{Finkelstein2022c,Harikane2022b,Bouwens2022b,Bouwens2022c,PerezGonzalez2023}. 
This excess is most pronounced in the brightest part of the luminosity function (LF), which seems to evolve very little from $z=4$ to $z=10$--12. This result is particularly tantalizing, because the brightest objects are detected at high significance in multiple bands and display a deep break, and are thus unlikely to be significantly contaminated by low-redshift interlopers \citep[e.g.,][]{Fujimoto2022}.

The origin of this excess is still to be determined --- it can be due either to a higher efficiency in star formation and galaxy assembly than previously thought, which makes them more abundant and/or massive, or to deviations from common prior assumptions on their physical properties [e.g., the initial mass function (IMF),  metallicity, and/or dust content] that increase their flux \citep[][]{Ferrara2022,Mason2022,Haslbauer2022,Kohandel2022,Ziparo2022,Fiore2022}.
More extreme explanations refer to nonstandard cosmological models to increase the abundance of bright and/or massive objects at very high redshift \citep[e.g.,][]{Melia2014,Melia2023,BoylanKolchin2022,Menci2022,Kannan2022}.

Among the various fields targeted by early {\it JWST} surveys, the first observations of a flanking field to the Hubble Frontier Field (HFF) cluster Abell 2744 \citep[A2744 hereafter;][]{Lotz2014,Castellano2016b} within the GLASS-JWST project delivered two of the brightest candidates at $z>10$, dubbed GHZ1/GLASSz10 (at $z\approx 10.5$) and  GHZ2/GLASSz12 (at $z\approx 12.3$) (\citeauthor{Castellano2022b} 2022a, C22a hereafter; \citealt{Naidu2022b,Bakx2022,Donnan2022,Harikane2022b,Yoon2022}). Compared to other detections (or lack thereof) in other and often larger fields, the GLASS-JWST field immediately qualified itself as one of the most interesting areas to investigate the most luminous sources at cosmic dawn.

Recently, more data have been obtained on the A2744 area by a number of {\it JWST} programs: the second epoch of the GLASS-JWST Early Release Science Program \citep[JWST-ERS-1324;][]{TreuGlass22}, the UNCOVER program \citep[PIs Labb\'e and Bezanson, JWST-GO\#2561;][]{Bezanson2022}, and the DDT program \#2756 (PI Wenlei Chen), which significantly extend and deepen the surveyed area. We take advantage of these new data and present here an analysis of the abundance of galaxies at $z\approx9$--11 selected in the NIRCam imaging data in the entire  A2744 region, with two goals: (i) improve the determination of the density of bright galaxies at $z\approx 10$, and (ii) start to characterize the clustering properties of this population.

Throughout the paper we adopt AB magnitudes \citep{Oke1983}, a \citet{Chabrier2003} IMF,   and a flat $\Lambda$CDM concordance model (H$_0$ = 70.0~km~s$^{-1}$ Mpc$^{-1}$, $\Omega_M=0.30$).

\section{NIRCam imaging and photometry} \label{sec:data}

\begin{deluxetable}{cccc}\label{tab:imaging}
\tablecaption{NIRCam Imaging}
\tablewidth{0pt}
\tablehead{\colhead{Filter} & \colhead{GLASS-JWST} & \colhead{DDT} & \colhead{UNCOVER}}%\\
\startdata
F090W & 29.5 & - & - \\
F115W & 29.7 & 28.2 & 29.1 \\
F150W & 29.5 & 28.3 & 29.0 \\
F200W & 29.6 & 28.6 & 29.0 \\
F277W & 29.8 & 28.8 & 29.3 \\
F356W & 29.9 & 28.8 & 29.4 \\
F410M & -    & -    &   28.8 \\
F444W & 29.6 & 28.6 & 28.9 \\
\enddata
\tablecomments{5$\sigma$ depths (mag) for point sources within a circular aperture of diameter $0\farcs2$.}
\end{deluxetable}

\begin{figure}[ht!]
\includegraphics[width=\linewidth,keepaspectratio]{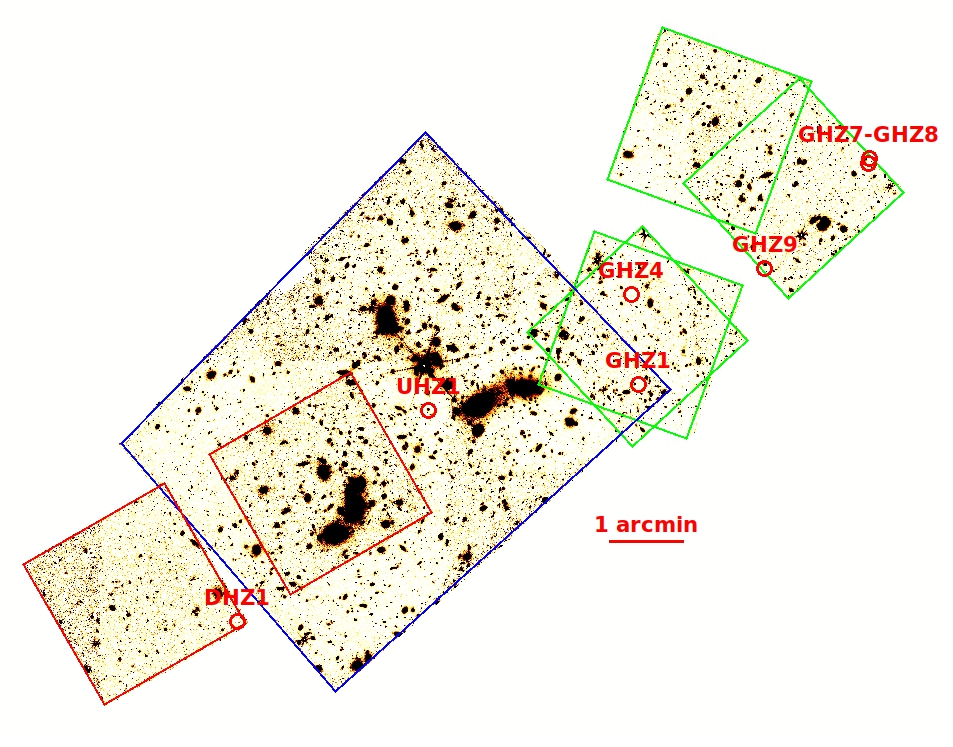}
\includegraphics[width=\linewidth,keepaspectratio]{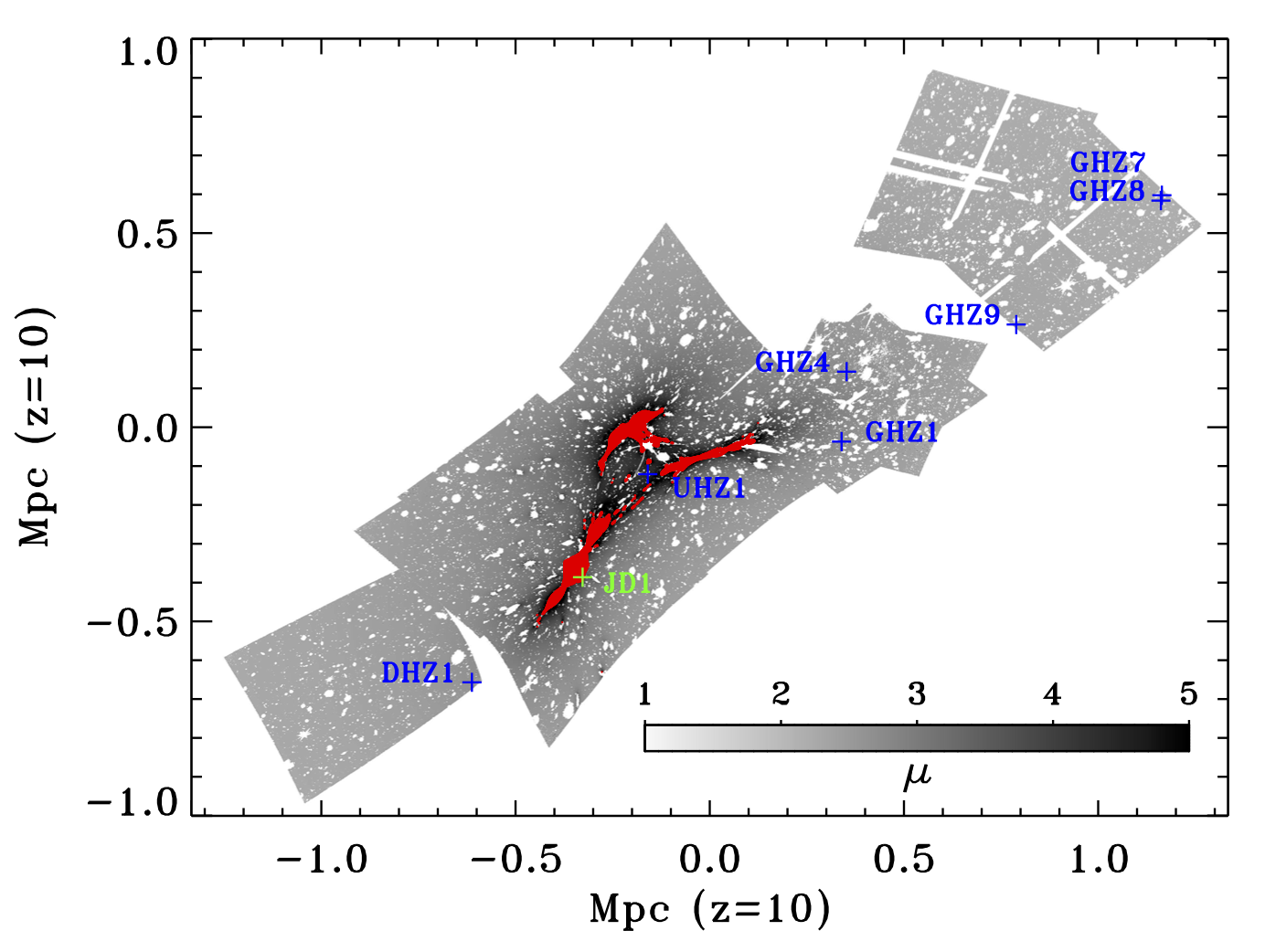}
\caption{\textbf{Top:} layout of the field and position of the $z\approx10$ candidates (red circles). The F444W mosaic combines observations from GLASS-JWST ERS (green region), UNCOVER (blue), and DDT-2756 (red). \textbf{Bottom:} the source plane magnification map at $z=10$. Masked regions outside the mosaics or covered by foreground objects are shown in white. The red region is the area with $\mu>5$ not used to compute the UV LF in the present paper; the triply lensed source JD1 at $z_{\rm spec}=9.76$ (green cross)  lies in this region and is not included in our sample. The source-plane positions of the $z\approx10$ candidates are marked with blue crosses.  \label{fig:field}}
\end{figure}

The NIRcam data analyzed in this paper are taken from three programs focused on the cluster A2744 and its surroundings. The GLASS-JWST NIRCam images have been taken in parallel to NIRISS on June 28--29, 2022, and in parallel to NIRSpec on Nov. 10--11, 2022. They consist of imaging through seven broad-band filters spanning from F090W to F444W as described by \citet{Treu2022}. The UNCOVER NIRCam observations of the cluster A2744 were taken on November 2, 4, 7, and 15, and exploit the same filter set as GLASS-JWST, except that the F090W filter is replaced with F410M. Finally, NIRCAM imaging of A2744 was obtained as part of DDT program 2756 on October 20.
The DDT-2756 (DDT, hereafter) filter set is the same as GLASS-JWST with the exception of the F090W filter, and overall shorter exposure times. The pointings overlap partially, resulting in the field geometry illustrated in Figure~\ref{fig:field} (top).
The image reduction, and the methods used to detect sources and measure multiband photometry, build on those described by \citet[][]{Merlin2022}, and take into account the improvements in data processing and calibration that have become available since then.

\begin{deluxetable*}{ccccccccc}\label{tab:candidates-photom}
\tablecaption{Photometry of the $z=9$--11 Galaxy Candidates in the GLASS-JWST, DDT, and UNCOVER Fields}
\tablewidth{0pt}
\tablehead{
\colhead{ID} & \colhead{F090W} & \colhead{F115W} & \colhead{F150W} &\colhead{F200W} & \colhead{F277W} & \colhead{F356W} & \colhead{F410M}& \colhead{F444W}}
%\decimalcolnumbers
\startdata          
GHZ1 & 2.4 $\pm$  2.3  & -2.1  $\pm$ 2.3 & 62.1 $\pm$  2.8 & 78.8  $\pm$ 2.5 & 78.7 $\pm$  2.2 & 81.7  $\pm$ 2.0 & -   &     111.8 $\pm$ 4.4    \\          
GHZ4 &-7.7  $\pm$ 4.9  &-4.4   $\pm$  4.9 & 29.9 $\pm$ 5.7  & 32.8 $\pm$ 5.3  &34.2 $\pm$  4.3 & 27.2 $\pm$  3.9 & -       &  41.7 $\pm$ 3.1      \\       
GHZ7 &-0.9 $\pm$ 2.4 &0.03  $\pm$  2.3  &32.4 $\pm$ 2.5  &38.2$\pm$  2.3 &30.9 $\pm$ 1.8 &25.9 $\pm$ 1.7 &-     &  28.8   $\pm$            3.6\\
GHZ8 &-1.3 $\pm$ 4.1 & -0.1 $\pm$ 3.9 & 51.9 $\pm$  4.4  &70.3 $\pm$  4.1 & 55.0 $\pm$  3.1 & 49.8  $\pm$ 2.9 & -  &   53.7  $\pm$            5.2      \\      
GHZ9 &3.3 $\pm$   2.2 & 2.2  $\pm$ 2.1 &29.5 $\pm$ 2.4  &24.2 $\pm$ 2.2 &31.2 $\pm$ 1.8 &35.8 $\pm$  1.6 &-        &  40.9  $\pm$            3.9    \\         
UHZ1 &-     &     4.1  $\pm$   6.5 &94.0 $\pm$  6.6 &100.6 $\pm$ 6.9  & 75.4  $\pm$ 4.5 & 65.6  $\pm$  3.6 &90.9$\pm$  6.7 & 87.7   $\pm$   7.2 \\      
DHZ1 &-  &    22.3  $\pm$                12.5      &        397.4   $\pm$           11.9      &        418.6     $\pm$        10.2          &    309.9     $\pm$         7.2      &         300.2       $\pm$       6.8         &     -   &     353.3     $\pm$         17.9 \\
\enddata
\tablecomments{Fluxes in nJy, not corrected for magnification.}
\end{deluxetable*}

A detailed description is given in a companion paper by \citet[][]{Paris2023}. For convenience of the reader, we briefly summarize below the information relevant for the present paper. Data reduction and flux calibration  were obtained using the official {\it JWST} pipeline\footnote{\url{https://jwst-docs.stsci.edu/jwst-science-calibration-pipeline-overview}} and exploiting the calibration files \textsc{jwst\_1019.pmap} made available by STScI in November 2022. We  then used a pipeline already adopted in similar projects \citep[]{Fontana2014}, based on \textsc{SCAMP} \citep[][]{Bertin2006} and \textsc{SWarp} \citep[][]{Bertin2002}, to combine the single exposures into mosaics projected onto a common  grid of pixels. For simplicity, considering the different depth and filter coverage of the three programs, and the need to perform simulations to assess the incompleteness in color space, we analyzed the three datasets independently and performed the simulations on each dataset separately. To avoid double counting and keep the fields independent from each other, certain regions were excluded from a given dataset; in practice, the UNCOVER NW corner was analyzed together with GLASS-JWST observations, and the portion of the DDT pointing that overlaps with UNCOVER was analyzed as part of the latter.

Objects were detected using a customized version of \textsc{SExtractor} \citep[][v2.8.6]{Bertin1996,Guo2013} on the F444W coadded images. Total F444W fluxes in elliptical apertures as defined by \citet[][]{Kron1980} were measured with \textsc{a-phot} \citep[][]{Merlin2019}. Fluxes in the other bands were measured with \textsc{a-phot} in several apertures at the positions of the detected sources using images point-spread-function (PSF)-matched to the F444W one. In the present analysis we use fluxes obtained by scaling the total F444W flux on the basis of the relevant color in $2 \times$ FWHM diameter ($= 0\farcs28$) apertures.

The mosaics are a combination of different exposures resulting in nonuniform depths. We summarize in Table \ref{tab:imaging} the typical 5$\sigma$ depths of the deepest regions for point sources within a circular aperture of diameter $0\farcs2$. Less-exposed regions can be 0.2--0.3 mag shallower depending on the band and field. The regions close to the cluster core are potentially affected by flux contamination from the intracluster light (ICL) \citep[e.g.,][]{Merlin2016}. We checked for possible systematics by injecting fake sources in different positions within the UNCOVER field.  We found no photometric offset, indicating that the ICL  contribution is effectively suppressed by the background-subtraction procedures adopted to build the final mosaics.  

We estimate total, nonoverlapping areas of 22.5, 10.2, and 4.6 sq. arcmin in the UNCOVER, GLASS-JWST, and DDT fields (respectively) available for the selection of high-redshift candidates by considering the regions observed in all bands, and excluding the pixels occupied by foreground sources.

\section{Selected galaxies at $z\approx10$}\label{sec:sample}
We select objects at $z\approx9$--11.5 using the color-color selection window defined by C22a:

\begin{equation*}\label{col1}
\begin{aligned}
    &(F115W-F150W)>1.7\\
    &(F115W-F150W)>2.17 \times (F200W-F277W)+1.7\\
    &-0.8<(F200W-F270W)<0.6
\end{aligned}
\end{equation*}

\begin{deluxetable*}{cccccccccccc}\label{tab:candidates}
\tablecaption{Properties of the Galaxy Candidates at $z=9$--11 in the GLASS-JWST, DDT, and UNCOVER Fields}
\tablewidth{0pt}
\tablehead{
\colhead{ID} & \colhead{R.A.} & \colhead{Dec} & \colhead{$M_{\rm UV}$} & \colhead{$z_{\rm zphot}$}  &  \colhead{$z_{\rm EAzY-v1p3}$} &  \colhead{$z_{\rm EAzY-Larson}$} & \colhead{SFR} & \colhead{$M_{\rm star}$} & \colhead{UV slope} &  \colhead{$\mu$}\\
\colhead{} & \colhead{deg.} & \colhead{deg.} & \colhead{mag} & \colhead{}  & \colhead{} & \colhead{} & \colhead{$M_{\odot}$ yr$^{-1}$} & \colhead{$10^8\,M_{\odot}$} &  \colhead{} & \colhead{}
}
\startdata
GHZ1 &  3.511929 & -30.371859  & -20.36$^{+0.31}_{-0.19}$  &  10.47$^{+0.38}_{-0.89}$ &  10.39$^{+0.19}_{-0.20}$ &  10.54$^{+0.20}_{-0.19}$ &  10.7$^{+42.7}_{-4.7}$  &  11.5$^{+7.1}_{-10.3}$  &-1.93$\pm$0.07&   1.71$^{+0.05}_{-0.05}$ \\
GHZ4 &   3.513739 & -30.351561 & -19.44$^{+0.16}_{-0.26}$  &  10.27$^{+1.2}_{-1.4}$ &  10.11$^{+0.46}_{-0.46}$ &  10.43$^{+0.61}_{-0.63}$ &   2.0$^{+14.2}_{-0.4}$ &  4.3$^{+1.5}_{-3.9}$& -2.31$\pm$0.36&   1.66$^{+0.05}_{-0.05}$  \\
GHZ7 &  3.451363 & -30.320718  & -20.06$^{+0.02}_{-0.17}$  &  10.62$^{+0.55}_{-1.02}$ &   9.97$^{+0.60}_{-0.32}$ &  10.57$^{+0.35}_{-0.33}$ &   3.2$^{+10.0}_{-0.5}$ &  2.1$^{+1.8}_{-1.7}$ & -2.66$\pm$0.15&   1.20$^{+0.01}_{-0.01}$ \\
GHZ8 &  3.451430 & -30.321796 & -20.73$^{+0.01}_{-0.01}$  &  10.85$^{+0.45}_{-0.57}$ &  10.14$^{+0.29}_{-0.28}$ &  10.79$^{+0.34}_{-0.34}$  &  17.5$^{+13.6}_{-12.3}$ &  0.8$^{+6.4}_{-0.16}$ &-2.60$\pm$0.14&   1.20$^{+0.02}_{-0.02} $ \\
GHZ9 &3.478756 & -30.345520  & -19.33$^{+0.04}_{-0.12}$  &   9.35$^{+0.77}_{-0.35}$ &   9.48$^{+0.40}_{-0.37}$ &   9.40$^{+0.20}_{-0.22}$ & 14.4$^{+15.0}_{-7.3}$ &  3.3$^{+2.9}_{-2.4}$ &-1.92$\pm$0.13&   1.33$^{+0.02}_{-0.02} $ \\
DHZ1  &  3.617257 & -30.425565  & -21.61$^{+0.03}_{-0.03}$  &   9.3127$\pm$0.0002$^a$  & &   &  25.4$^{+3.2}_{-4.3}$&  25$^{+6.6}_{-5.0}$ &-1.80$\pm$0.08 &   1.66$^{+0.02}_{-0.01} $ \\
UHZ1 &  3.567065 & -30.377857 & -19.79$^{+0.16}_{-0.17}$  &  10.32$^{+0.25}_{-1.0}$ &  9.88$^{+0.21}_{-0.19}$ &  9.99$^{+0.47}_{-0.48}$ &   4.5$^{+2.9}_{-2.2}$ &  0.4$^{+1.8}_{-0.2}$ &-2.72$\pm$0.15&   3.72$^{+0.23}_{-0.24} $ \\
\enddata
\tablecomments{The demagnified, rest-frame $M_{\rm UV}$ have been obtained at the best-fit \textsc{zphot} redshift, and the uncertainties include the contribution of both photometry and magnification. Stellar masses and SFRs have been obtained at the best-fit \textsc{zphot} redshift as in \citet[][]{Santini2022} and corrected for magnification. Uncertainties include error contribution from SED fitting and magnification. The UV slope $\beta$ is measured by fitting the F200W, F277W, and F356W bands; the uncertainties in the fit account for photometric errors \citep[][]{Castellano2012}. The uncertainties in the magnification $\mu$ are at the 68\% confidence level; see Sect.~\ref{sec:lensmodel}. $^a$Spectroscopic redshift  from \citet[]{Boyett2023}. All properties of DHZ1 have been measured fixing the redshift at the spectroscopic value.}
\end{deluxetable*}

\begin{figure*}[ht!]
\includegraphics[width=9cm]{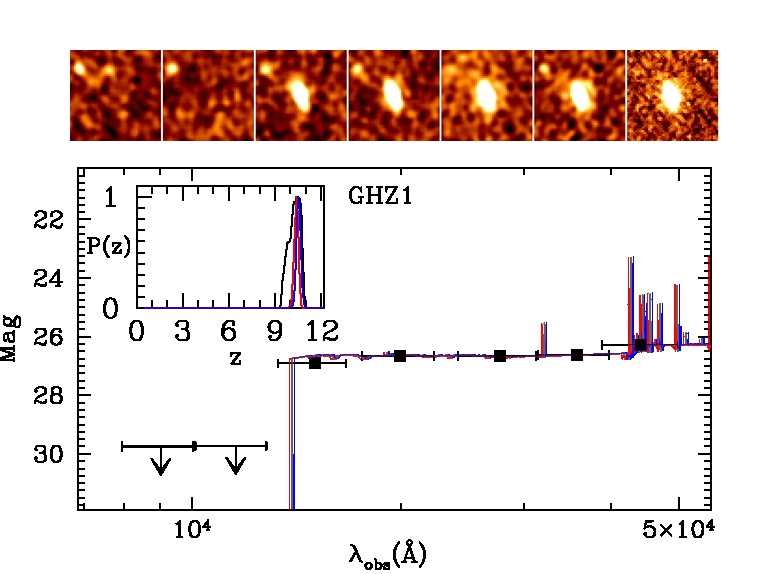}
\includegraphics[width=9cm]{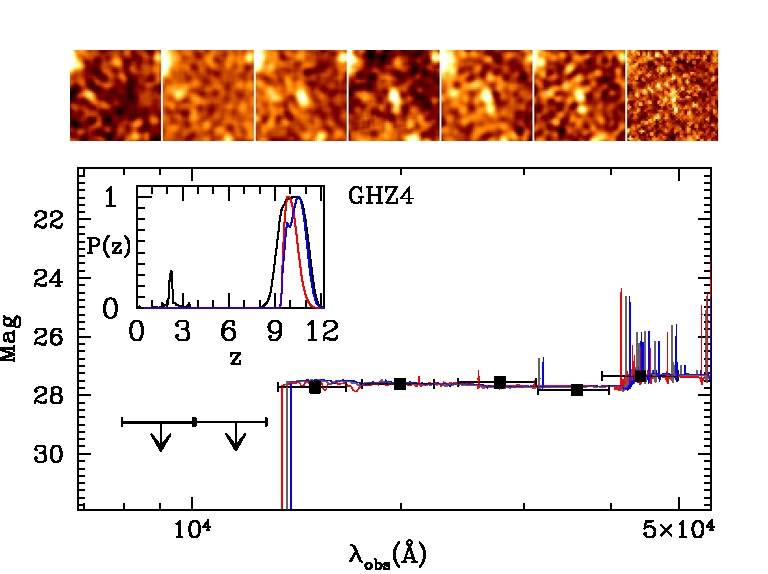}
\includegraphics[width=9cm]{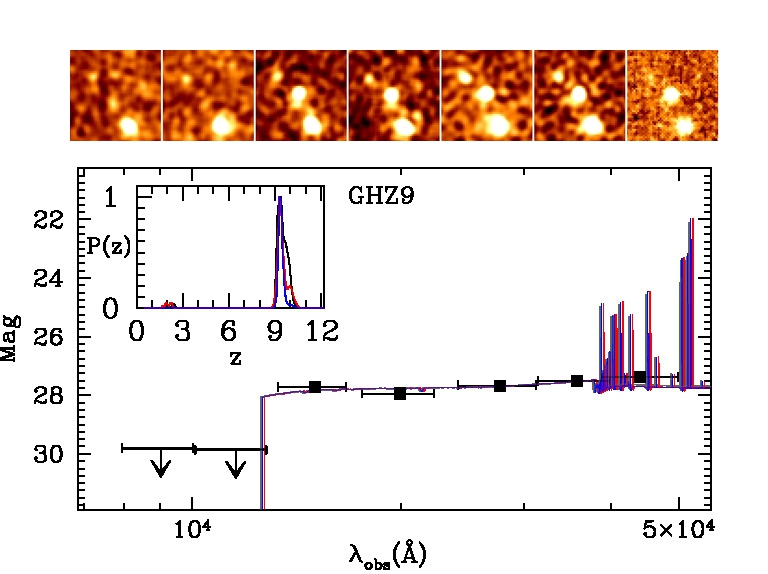}
\includegraphics[width=9cm]{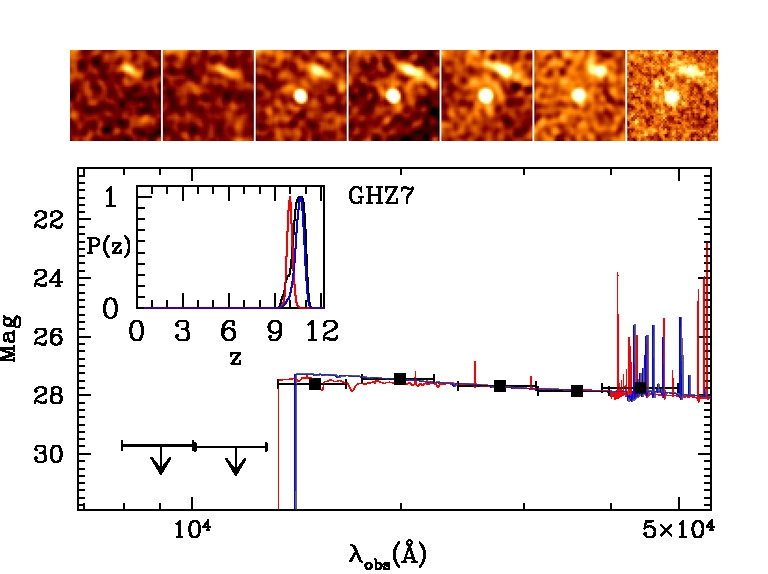}
\includegraphics[width=9cm]{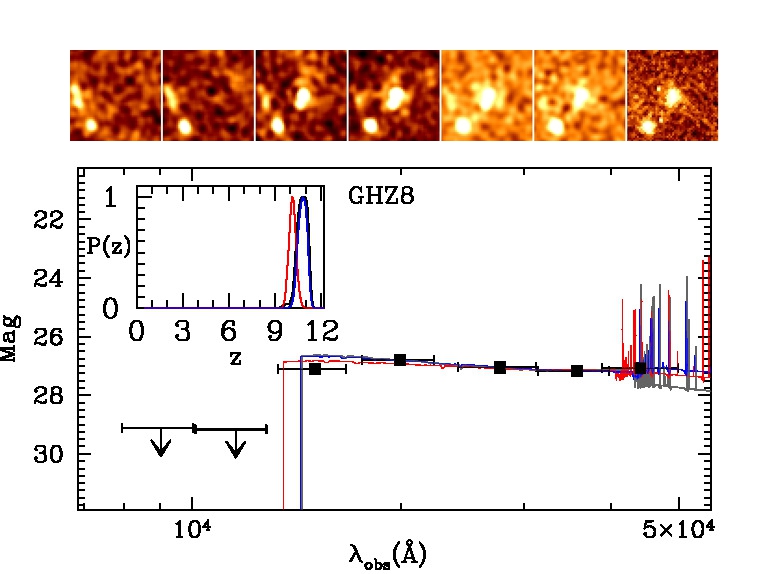}
\caption{The $z\approx10$ candidates from the GLASS-JWST field. Photometry (not corrected for magnification) and SEDs are given in the main quadrant. Upper limits are reported at the $2\sigma$ level. We show in the inset the redshift probability distributions $P(z)$ from \textsc{zphot} (gray) and \textsc{EAzY} (red for the standard V1.3 templates, and blue for the \citealt{Larson2022} templates). The SEDs are obtained by fitting the BC03 library described in Sect.~\ref{sec:sample} at the best-fit redshifts from the relevant \textsc{zphot} and  \textsc{EAzY} runs.  Thumbnails, from left to right, respectively show the objects in the F090W, F115W, F150W, F200W, F277W, F356W, and F444W bands.\label{fig:SEDs-GLASS}}
\end{figure*}

\begin{figure}[ht!]
\includegraphics[width=9cm]{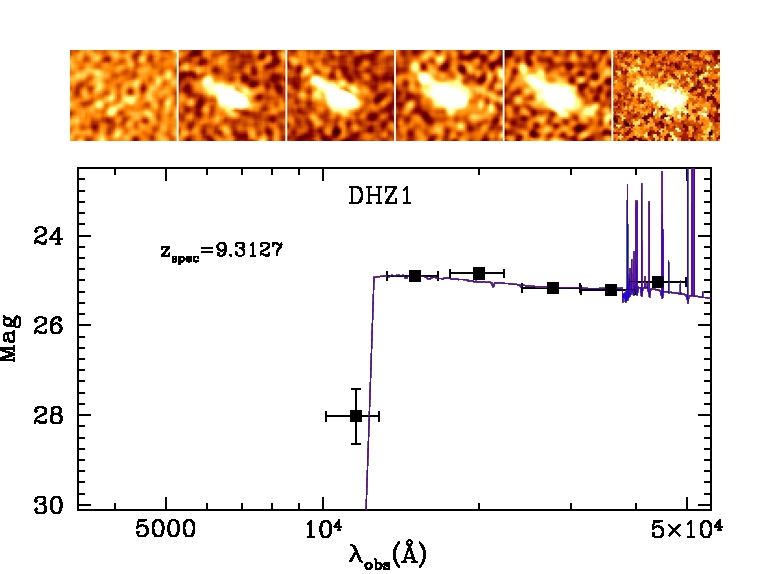}
\includegraphics[width=9cm]{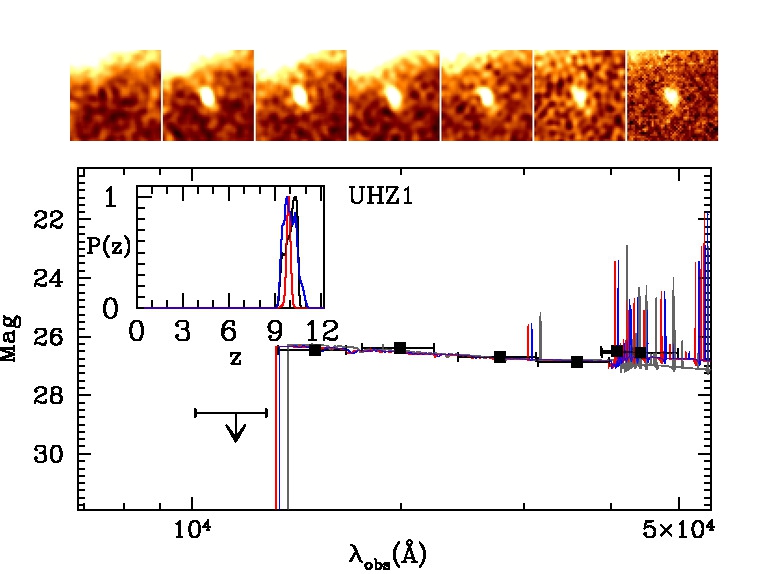}
\caption{Same as Fig.~\ref{fig:SEDs-GLASS} for the $z\approx10$ candidates in the DDT field and in the low-magnification area of the UNCOVER field. Thumbnails, from left to right, respectively show the objects in the F115W, F150W, F200W, F277W, F356W, F410M (UHZ1 only), and F444W bands.\label{fig:SEDs-DDTUNCOVER}}
\end{figure}

\noindent
We require signal-to-noise ratio (SNR) $>8$ in the F444W band, SNR $>2$ in the other bands redward of the Lyman break, and SNR $<2.0$ in the F090W band, where available. 
The selection yields 5 candidates in the GLASS-JWST field, 2 in UNCOVER, and 1 in the DDT region. The JWST photometry of the candidates is listed in Table~\ref{tab:candidates-photom}.

The five selected objects in GLASS-JWST include GHZ1 and GHZ4 from C22a, and three galaxies in the region recently observed in the NIRSpec parallel and discussed here for the first time. In UNCOVER we select a new robust candidate UHZ1 with $m_{F444W}=26.54\pm0.09$ mag, outside of the area previously observed by the HFF. We note that object JD1B from \citet[][]{Zitrin2014}, recently confirmed to be at $z=9.76$ with NIRSpec \citep[][]{RobertsBorsani2022d}, is also formally selected by our criteria. However, we will not further discuss object JD1B, one of the multiple images of a strongly magnified ultrafaint galaxy, and it will not be used to estimate the ultraviolet (UV) LF in the present analysis which is limited to moderate-magnification regions (see Sect. 4). We checked the other counterimages reported by \citet[][]{Zitrin2014}, finding that JD1A is just outside the color-selection region, while the faintest one (JD1C) is not detected likely owing to the strong background in the F444W band from a nearby cluster galaxy. The nondetection of JD1A and in particular of JD1C are not surprising, considering the selection completeness level in the UNCOVER field (see Sect. 5) and that the catalog extraction is not optimized for the analysis of the inner cluster regions affected by significant crowding. The only candidate selected in the DDT program is also the brightest in the sample, with $m_{F200W}=24.84\pm0.05$ mag. 

Objects GHZ1, GHZ4, and GHZ9 (from GLASS-JWST), and the UNCOVER and DDT candidates, have also been observed with {\it HST} under the HFF and BUFFALO \citep[][]{Steinhardt2020} programs. We further assessed their reliability by measuring the SNR at their position in the F606W, F814W, and F105W mosaics processed with the \texttt{grizli} pipeline \citep[][]{Brammer2022} in 0\farcs2 diameter apertures, finding SNR $<2$ in all cases.

For all selected candidates we estimate photometric redshifts with \textsc{EAzY} \citep{Brammer2008} and \textsc{zphot} \citep{Fontana2000}. The \textsc{EAzY} code was run assuming a flat prior with two different sets of templates: the default V1.3 spectral template \citep[see][]{Leethochawalit2022}, and the recently released templates by \citet{Larson2022} specifically designed for the analysis of {\it JWST}-selected galaxies at very high redshift. The analysis with \textsc{zphot} has been performed as described by \citet[][]{Santini2022} by fitting the observed photometry with \citet[][BC03 hereafter]{Bruzual2003} templates having both declining and delayed star-formation histories; it models the contribution from nebular continuum and line emission following \citet{Schaerer2009} and \citet{Castellano2014}. We show spectral energy distributions (SEDs), photometric redshift probability distributions, and thumbnails of the seven candidates in Figs.~\ref{fig:SEDs-GLASS} and  \ref{fig:SEDs-DDTUNCOVER}. Their properties are listed in Table~\ref{tab:candidates}, including demagnified $M_{UV}$ and $\mu$ values based on the lensing model that will be discussed in Sect.~\ref{sec:lensmodel}. 

The two objects GHZ1 and GHZ4 from C22a fall in the region recently reobserved in parallel to NIRSpec. Reassuringly, their SEDs and photometric redshifts obtained from the new data are in excellent agreement with those measured in the first GLASS-JWST NIRCam observations, with the deeper data analyzed here reducing the likelihood of the secondary photometric redshift solution at $z\approx2$ for GHZ4. Candidate GHZ1 also falls in the area that we have removed from the UNCOVER field to avoid duplications. The availability of the lensing model extending to the GLASS-JWST region allows us to estimate that they are moderately magnified, such that the intrinsic $M_{UV}$ are $\sim0.5$--0.6 mag fainter than the estimates reported in our previous paper. The three newly found  candidates in the newly observed GLASS-JWST area have observed $m_{F444W}\approx27$--27.3 mag and similar SEDs. Two of them (GHZ7 and GHZ8, see Fig.~\ref{fig:field}) have very similar $P(z)$ and are separated by only $4''$ in the sky, corresponding to a projected separation of 16~kpc at $z\approx10$ --- i.e. they are likely to be physically associated in a close pair. Two of the other three candidates previously reported by C22a (GHZ5 and GHZ6) do not enter the present sample, as with the new photometry they are located just outside the color-selection region. Their photometric redshift solutions still show a significant peak at $z>9$ as reported by C22a, but they are not included in the present analysis which is based only on color-selected candidates. A third candidate (GHZ3) of the C22a paper is instead detected in the new and deeper images obtained in the F090W filter, and it is therefore definitely not a $z\approx 10$ galaxy. These findings are consistent with both the effect of photometric scatter, and, as far as GHZ3 is concerned, with the possible contamination level discussed by C22a. However, at variance with the C22a sample, none of the candidates discussed in the present paper show significant secondary photometric redshift solutions, most likely owing to the availability of deeper data in the GLASS-JWST field. 

We note that various candidates are found not far from the image edges. However, they all fall in full-depth regions of the mosaics and we made sure that their detection and photometry are not affected by border effects. The most remarkable object in the present sample is galaxy DHZ1 from the DDT area. This object had also been previously selected as a high-redshift candidate from HFF imaging \citep[][]{Castellano2016b,Yue2018}, and it has been recently spectroscopically confirmed at $z=9.3127$ with NIRSPec \citep[][]{Boyett2023}. DHZ1 has a demagnified $M_{\rm UV}=-21.6$~mag, and a complex morphology indicative of an ongoing interaction or merging comprising two main components and an elongated tidal tail. In the following we will adopt for DHZ1 the redshift and parameters from \citet[][]{Boyett2023}. 

Basic physical properties of the sample galaxies are summarized in Table~\ref{tab:candidates}. Corrected for lensing magnification, the star-formation rates (SFRs) range between 2 and 25~$M_{\odot}$~yr$^{-1}$, the estimated stellar masses range between $4\times10^7$ and  $1.1\times10^9$~$M_{\odot}$, and the UV slopes $\beta$ range between $-1.8$ and $-2.7$. We note that at $z\geq 10$  the rest-frame optical bands are poorly constrained by NIRCAM data, at best in regions that can be contaminated by emission lines. Consequently, stellar masses can be subject to uncertainties and biases, and stellar ages are  poorly constrained. For this reason we do not quote the latter in Table~\ref{tab:candidates}. The reported values appear to encompass the range of rest-frame properties of Lyman-break galaxies at these redshifts, as observed by recent surveys. We caution that these estimates have been obtained under the assumption of a \citet{Chabrier2003} IMF, but may be substantially different if the underlying galaxy-wide IMF \citep[e.g.,][]{Yan2021} is top-heavy, as proposed to explain the unexpected abundance of bright high-redshift galaxies in early {\it JWST} observations \citep[][]{Haslbauer2022,Finkelstein2022c}. Nonetheless, we report the estimated values to facilitate follow-up observations and we defer a more detailed analysis of the properties of these galaxies to future work. 

\section{Methodology}
\label{secmethods}

Estimating the LF of high-redshift galaxies in lensed fields is more complex than in blank fields. The additional complexity is usually justified by the extra gain in depth afforded by lensing magnification.  In this work, however, motivated by the discovery of a significant number of bright $z\approx 10$ galaxies, we focus primarily on the bright end of the LF, and postpone the study of the faintest objects to a future work. This choice allows us to keep the analysis simple and robust, building on the ingredients that we discuss below.

\subsection{Lens Model}
\label{sec:lensmodel}

The A2744 region, especially that covered by the UNCOVER data, is affected by moderate to strong lensing amplification due to the foreground cluster, whose mass distribution is known to extend over a wide area covering the entire dataset \citep[e.g.,][]{Merten2011}. To properly account for the lensing effect, we have extended the pre-{\it JWST} lens model described by \citet[][\citetalias{Bergamini2022} hereafter]{Bergamini2022} by including 27 additional multiple images in the cluster core, as well as 29 in the NW region where two cluster galaxies (named G1 and G2 by \citetalias{Bergamini2022}) are present with luminosities similar to those of the central bright cluster galaxies. Owing to the rich strong-lensing (SL) features revealed by the NIRCam data, we refer to this region as ``SL clump." The identification of extra multiple images takes advantage of NIRCam multiband photometry, as well as new spectroscopic observations with VLT/MUSE \citep{PrietoLyon2022, Bergamini2023}. MUSE coverage of the SL clump also enables the identification of additional cluster galaxies and the determination of their
velocity dispersions, which are found to be consistent with the subhalo scaling relation resulting from 
the lens-model optimization. The total root-mean-square (RMS) separation between the observed and model-predicted positions of the 149 multiple images is $0\farcs43$. The extended and enhanced SL model is described in detail by \citet{Bergamini2023}.

For the purpose of the present work, we have computed the median magnification values and 95\% confidence level intervals for each high-$z$ candidate by extracting 100 random sets of parameter values from the final Markov chain Monte Carlo (MCMC) chain, containing a total of $3\times 10^4$ samples. To account for systematic uncertainties owing to the choice of the total mass parametrization in the external region, we have combined the MCMC chains of two extended lens models, which include either one or two cluster-scale halos in the SL clump, using the same set of multiple image constraints.

To derive the LF of the high-$z$ candidates, unlensed survey areas and volumes in the considered redshift bins are needed. Intrinsic survey areas are computed for each of the three fields by projecting the footprints of the observed regions on the source plane, at varying redshift in the range $9.0<z<11.5$. To this aim, we use the deflection field from the extended lens model with the best positional RMS uncertainty (i.e., the one including only a single halo in the SL clump region). We also project onto the source plane the observed mask (excluding regions around bright sources) and the magnification maps in the aforementioned redshift range (see Figure~\ref{fig:field}-bottom). In this way, we can reconstruct the effective area on the source plane, for each of the three fields separately, where objects can potentially be identified around $z=10$, and compute the corresponding magnification in each pixel. These maps are then combined to estimate the probed survey volume in each redshift and luminosity bin.

The seven galaxies at $z\approx 10$ listed in Table~\ref{tab:candidates} have low-to-moderate magnifications, ranging from 1.2 to $\sim 4$ (see Table~\ref{tab:candidates}), and are not multiply imaged (i.e., they lie outside the main caustics on the source plane at $z\approx 10$). This allows us to limit our LF analysis
to regions with $\mu<5$, which has the significant advantage of avoiding strongly lensed regions where systematic uncertainties can be large, albeit associated with small survey volumes. As previously mentioned, JD1B is the only object at $\mu>5$ meeting our color-selection criteria and is thus excluded from the sample analyzed in this paper. The corresponding areas with $\mu<5$ in the source plane amount to 6.86, 9.44, and 3.26 sq. arcmin in the GLASS--JWST, UNCOVER, and DDT fields, respectively.
By comparing the effective source plane area predicted by lens models with slightly different mass parametrizations, but similar positional RMS values, we estimate the systematic uncertainty on the effective survey volume (and therefore on the data points in the LF) to be approximately 5\%.

We have also checked the predicted magnification values for our objects by the SL model recently released by the UNCOVER collaboration \citep{Furtak2022}\footnote{Magnifcation maps made public as of Dec. 12, 2022.}. Four of our seven candidates have magnification values at $z\approx10$ lower than 5 and in good agreement with the ones presented here, while the remaining three (i.e., GHZ7, GHZ8, and GHZ9) are not covered by their model.

\begin{deluxetable}{ccc}\label{tab:UVLF}
\tablecaption{Binned Luminosity Function}
\tablewidth{0pt}
\tablehead{
\colhead{$M_{\rm UV}$} & \colhead{$N_{\rm obj}$} & \colhead{$\phi$} \\ 
\colhead{} & \colhead{} & \colhead{$10^{-5}$ Mpc$^{-3}$ mag$^{-1}$}}
\startdata
$-22.5 \pm 0.5$ & 0 & $<$3.3\\ %3.3e-05
$-21.5 \pm 0.5$ & 1 & 2.1$^{+4.8}_{-1.7}$\\
$-20.5 \pm 0.5$ & 3 & 7.6$^{+6.4}_{-3.9}$\\
$-19.5 \pm 0.5$ & 3 & 18.0$^{+17.5}_{-9.8}$\\
\enddata
\tablecomments{The binned volume densities in the GLASS-JWST, UNCOVER, and DDT fields; uncertainties are computed for small-number Poisson statistics following \citet[][]{Gehrels1986}.}
\end{deluxetable}

\begin{figure}[ht!]
\includegraphics[trim={1.0cm 1.5cm 0.3cm 0.3cm},clip,width=\linewidth,keepaspectratio]{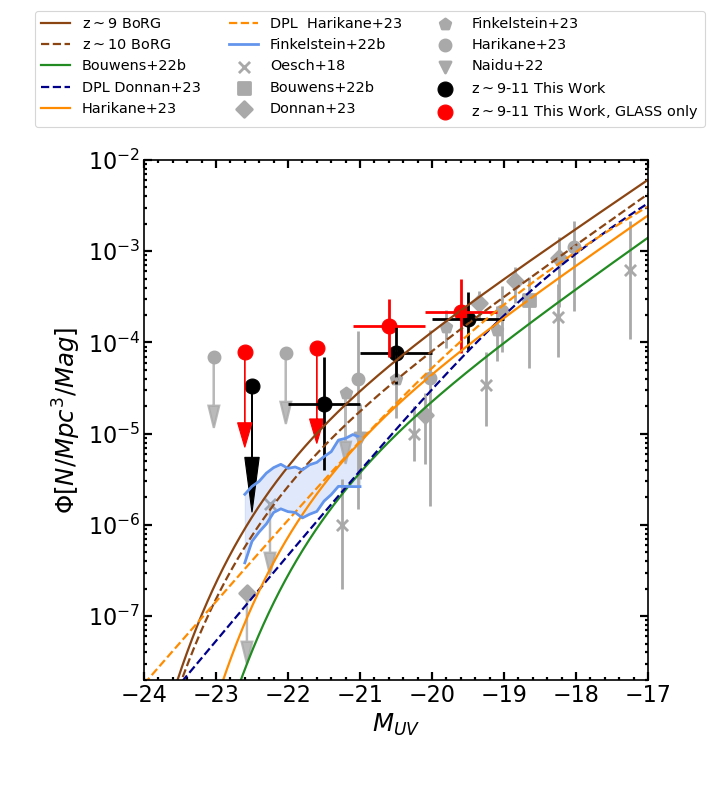}
\caption{The UV LF at $z\approx10$ in the A2744 region observed by the GLASS-JWST, UNCOVER, and DDT\#2756 programs. Black circles and error bars are obtained from the full sample, red ones (shifted by 0.1 mag for clarity) for the GLASS-JWST field only. The uncertainties are given by small-number Poisson statistics in each bin following \citet[][]{Gehrels1986}. Binned LFs from the literature are shown as gray symbols (see label for details). The shaded blue region indicate the constraints by \citet[][]{Finkelstein2022} from HST-CANDELS.  We also show the LF analytic fits estimated by \citet[][BoRG survey $z\approx9$--10]{Leethochawalit2022a}, \citet[][]{Bouwens2022b}, \citet[][]{Donnan2022}, and \citet[][]{Harikane2022b}.\label{fig:UVLF}}
\end{figure}

\section{The $z\approx10$ UV Luminosity Function in the A2744 Region} \label{sec:results}

Even correcting for magnifications, our sample has remarkably bright UV magnitudes $M_{\rm UV}\lesssim -20$, corresponding to a high surface density of objects. We quantify the volume density by computing the UV LF in four evenly spaced rest-frame magnitude bins at $-23.0 \leq M_{\rm UV}\leq -19.0$~mag.  
The effective volumes in each bin are obtained by taking into account the area at different magnification levels, and the relevant completeness for the selection of objects with the considered UV rest-frame magnitudes. In fact, the volume accessible for sources of a given \textit{intrinsic} magnitude $M_{\rm UV,int}$ is
\begin{equation}
V_{{\rm eff},M_{\rm UV,int}}=\int_{z=9}^{z=12} \int_{\mu=1}^{\mu=5} C(M_{\rm UV,obs},z)\frac{dV(\mu,z)}{dz} \,d\mu  dz\, ,
\end{equation}

\noindent
where $M_{\rm UV,obs}=M_{\rm UV,int}-2.5\,{\rm log}(\mu)$. In practice, the effective volumes are calculated for the three bins reported in Table~\ref{tab:UVLF} by replacing the integral with sums in steps of $\Delta\mu=0.25$ and $\Delta z=0.25$. The completeness values $C$ at different redshifts and magnifications are obtained on the basis of  imaging simulations, and the element volumes $dV(\mu,z)$ available for the selection of high-redshift galaxies are computed as a function of $\mu$ on the basis of the area of the masked magnification maps. 

The choice of limiting the analysis to galaxies identified in low-magnification areas, while not significantly reducing the probed volume, also simplifies the estimate of the completeness which does not require simulating multiply imaged objects. Imaging simulations are performed separately for the GLASS-JWST, UNCOVER, and DDT fields as described by C22a. Briefly, we inserted in blank regions of the observed images $3 \times 10^5$ mock Lyman-break galaxies (LBGs) at $9<z<12$ and with a uniform distribution at $-18.5 < M_{\rm UV}< -23$ mag. The observed magnitudes are obtained by randomly associating a model from  a library based on BC03 models with metallicity $Z =0.02\,Z_{\odot}$, $0 < E(B-V) < 0.2$ mag, constant star-formation history, \citet{Salpeter1955} IMF, and \citet{Calzetti2000} extinction law. We assume that objects follow a circular \citet{Sersic1968} light profile with index $n=1$ and that their effective radius scales with $L_{\rm UV}$ as $r_{e} \propto L^{0.5}$, consistent with several estimates at comparable redshifts \citep[e.g.,][]{Grazian2012,Kawamata2018,Bouwens2022,Yang2022}. Following \citet{Yang2022}, we assume an effective radius of 0.8 kpc for objects with $M_{\rm UV}=-21$ mag. In order to avoid overcrowding, simulations are repeated by inserting 500 objects each time. Detection, photometry, and color selection on the simulated galaxies are performed in the same way as for the real catalogs. 

We consider for each object the demagnified $M_{\rm UV}$ reported in Table~\ref{tab:candidates} which has been computed on the basis of the best-fit \textsc{zphot} template. Considering the width of the magnitude bins, magnification and redshift uncertainties do not significantly affect our results. 
The binned volume densities are then obtained as $\phi_i=N_i/V_{{\rm eff},i}$, where $N_i$ is the number of objects in the considered bin. The imaging simulations confirm that our selection criteria isolate objects at $9 \lesssim z \lesssim 11.5$. The peak in the selection efficiency is found at $z\approx 10$, where the completeness $C$ is higher than 90\% for $M_{\rm UV}\approx -20$~mag objects in the GLASS-JWST field --- the deepest of the three --- while it is $\sim$40\%--80\% in DDT and UNCOVER (respectively)  because of shallower depths and, in the case of UNCOVER, crowding. As a result, objects in the faintest magnitude bin can only effectively be selected in UNCOVER and GLASS-JWST ($C\approx 60$--80\%), while the DDT field provides only a minor contribution to the effective volume ($C\approx 20$\%).

The binned UV LF is shown in Fig.~\ref{fig:UVLF}, together with results from recent surveys in the same redshift range. The relevant values are presented in Table~\ref{tab:UVLF}. We compute and plot in Fig.~\ref{fig:UVLF} the LF derived for the entire sample and considering GLASS-JWST only.
It is immediately clear that the density of $z\approx 10$ sources in our sample is significantly larger than the best-fit LF derived from previous {\it JWST} surveys --- which in most cases include the first epoch of the GLASS-JWST data, and so may already be  biased high.  The density estimated from our entire sample is higher, but still consistent at $1\sigma$, with that inferred at $z\approx10$ by previous wide-field imaging surveys with {\it HST}, such as BoRG \citep{Leethochawalit2022a,Bagley2022}, that do not include the GLASS-JWST data and were selected with a different filter complement. The discrepancy is particularly remarkable at $M_{\rm UV}\approx -21$~mag, where it ranges between 4 and 10 times with respect to previous estimates, although of course there is considerable uncertainty at this stage. On the basis of the number of objects and on the surveyed volume, we estimate a  cosmic variance at the level of $\sim13$--15\%\footnote{Following the Cosmic Variance Calculator at \url{https://www.ph.unimelb.edu.au/~mtrenti/cvc/CosmicVariance.html}. See \citet[][]{Trenti2008}.} in each of the three fields, where the highest value refers to the DDT region. This is much smaller than the Poisson uncertainty in the number counts.

While the high density measured in the field is suggestive of the presence of a physical overdensity, the accuracy of photometric redshifts prevents to confirm whether the objects, or part of them, are members of a localized structure. In this respect, we note that GHZ9 has best-fit photometric redshift solutions at $z \simeq 9.3-9.5$, compatible with the spectroscopic value measured for DHZ1, while objects GHZ1, GHZ7 and GHZ8 are more likely placed at $10 \lesssim z \lesssim 11.5$. In turn, GHZ4 and UHZ1 are compatible with both the $z\sim 9-10$ and the $z \sim 10-11$ redshift ranges. As a test, we estimated the UV LF in the two contiguous redshift bins at $9 \leq z \leq 10$ (4 objects, including GHZ4 and UHZ1) and $10 < z \leq 11.5$ (3 objects). We find that the number density of $M_{\rm UV}\approx -20.5$~mag objects at $z\approx10-11$ in the A2744 field is higher than previous estimates, including the $z\sim 10$ BoRG one but still consistent at $1\sigma$ with the latter. The UV LF at $z\approx9-10$ is in agreement with the BoRG $z\sim 9$ estimate at intermediate magnitudes, but DHZ1 remains unexpected considering the probed volume as it points to a number density of extremely bright objects $>$3 times higher than predicted. Summarising, even when conservatively dividing the sample on the basis of the photometric redshifts, the estimated UV LFs point to a higher-than-expected density in the A2744 region.

Interestingly, we find an even higher density if we restrict the analysis to the GLASS-JWST field (red points in Fig.~\ref{fig:UVLF}). This  is  the deepest area considered, and the region least affected by magnification ($\mu\approx1.2$--1.7), so it is difficult to attribute this result to a higher contamination. In this region, our inferred density is higher even than the $z\approx10$ BoRG estimates at $M_{\rm UV}\approx -20.5$~mag. Furthermore, we tested that when restricting the analysis to objects with photometric redshift in the interval $10<z<11$ in this field, our estimated density at that magnitude further increases by a factor of $\sim 2$. We thus suggest that the GLASS-JWST area possibly includes an overdensity.  The sources appear aligned along a filament-like structure of projected length $\sim 2$~Mpc with the pair GHZ7--GHZ8, separated by 16 kpc. The galaxies have similar colors and $P(z)$, resembling similar galaxy associations found in other high-redshift structures \citep[e.g.,][]{Castellano2018,Castellano2022a}.

The sample size is too small to carry out a meaningful study of the two-point correlation function, but determining the clustering properties of bright galaxies at $z\approx10$ will be important to determine their average cosmic density, their halo mass, and their role in cosmic reionization \citep[][]{Endsley2020}. As a first assessment, we have used the $M_{\rm UV}- M_{h}$ relation between UV magnitude and halo mass at $z\approx10$ from \citet[]{Mason2022} to estimate a lower limit to the total halo mass of the potential overdensity as the sum of the masses of the objects selected here. We find $M_{h}=6.7 \times 10^{11}~M_{\odot}$ when considering the entire sample, and $M_{h}=3.4 \times 10^{11}~ M_{\odot}$ under the assumption that the overdensity is localized in the GLASS-JWST region. We then analysed 10 (26) independent lines of sight with the same area as the A2744 (GLASS-JWST) region from a light cone extracted with the \texttt{FORECAST} software (Fortuni et al., in prep.) from the IllustrisTNG simulation \citep{Pillepich2018}. We found no halos as massive as the ones estimated in our case, indicating a  probability smaller than 5--10\% of finding such an overdensity in our survey.

To obtain a more quantitative assessment, we estimated the rarity of the potential overdensity starting from the computation of the linear RMS density fluctuation $\sigma$ predicted by the standard $\Lambda$CDM cosmology.
The cosmic volumes we consider correspond to mass scales
$M_{\rm vol}=3.4 \times 10^{12}~M_{\odot}$ and $M_{\rm vol}=1.16 \times 10^{12}~M_{\odot}$ on the entire field and in the GLASS-JWST field, respectively. Assuming a $\Lambda$CDM power spectrum, the above mass scales correspond (at $z\approx 10$) to $\sigma=0.25$ and $\sigma=0.3$, respectively.
On the physical scales of our system, the density field is nonlinear, and a possible approximation for the true density distribution $P(\delta,\sigma,\mu_s)$ is a log-normal model
\citep[]{Klypin2018} for which $\sigma$ can be approximated as the RMS in $ln(\delta)$ with a mean in this quantity of $\mu_s=-\sigma^2/2$. Under these assumptions, the probability of finding an overdensity $\delta \approx 4$ or larger in the entire volume sampled by the A2744 region is $P(\geq \delta|\sigma,\mu_s)=0.09$, while the probability of finding an overdensity $\delta \approx 8$ or larger in the volume sampled by the GLASS-JWST field is $P(\geq \delta|\sigma,\mu_s)=0.01$.

These tests show that the potential overdensity in the A2744 region is rare, in particular as far as the high density contrast measured in the GLASS-JWST area is concerned, and unlikely to be found in this survey, also considering that the region  hosts another rare overdensity at a slightly lower redshift \citep[][]{Morishita2022}. Clearly, the accuracy achieved by photometric redshifts does not allow us to firmly constrain the presence of a localized structure. Extensive follow-up spectroscopy of the A2744 region is needed to confirm this hypothesis, and measure the redshift and extent of the possible overdensity.

\section{Summary and Conclusions} \label{sec:summary}

The rapid discovery of an unexpectedly large number of galaxies at $z\approx 10$ and beyond is one of the most tantalizing results obtained by the very first {\it JWST} observations. The higher than expected normalization of the UV LF at  $z>9$ \citep[e.g., C22a;][]{Finkelstein2022c} has already generated a lively theoretical debate about its potential physical explanations \citep[e.g.,][]{Mason2022,Ferrara2022}. In this context, it is important to carry out further studies to improve the statistics and start to explore the all-important issues of clustering and cosmic variance.

We report here the results of a search for $z\approx 10$ galaxies in a new set of {\it JWST} imaging data, obtained by three different programs (GLASS-JWST, UNCOVER, and DDT\#2756) in the region of the cluster Abell 2744. The combination of depth (GLASS-JWST is among the deepest fields obtained so far with {\it JWST}) and lensing magnification (although we limit our analysis to $\mu<5$, where the lens model is robust) allows us to select with high reliability galaxies down to an intrinsic rest-frame luminosity of $M_{\rm UV}\approx -19$ mag.  Remarkably, we identify 7 galaxies at this redshift, including two previously reported in the first epoch of the GLASS-JWST dataset, and one object with spectroscopic confirmation at $z=9.3127$. Five of the galaxies are  detected in the (relatively small) GLASS-JWST area, and are distributed along a filament-like structure $\sim2$ Mpc in projected length. This sample consists of an excellent set of candidates for efficient follow-up spectroscopy with NIRSpec.

Building upon a revised lensing model that improves the description of the entire A2744 area by taking advantage of new multiple images and MUSE spectra, we compute the resulting number density in three magnitude bins and compare it with the results from recent {\it JWST} surveys. 
We conclude that the density of bright $z\approx 10$ galaxies in the A2744 region is significantly higher than average, by factors that range between 3 and 10 depending on $M_{\rm UV}$  and on the survey taken as reference.

We conclude that extending the search in a larger area northwest of the cluster is necessary to further characterize this potential overdensity. Spectroscopic redshifts can determine whether these objects are part of a single physical overdensity \citep[e.g.,][]{Laporte2022}, or the apparent density is enhanced by chance superposition of galaxies at similar redshifts \cite[e.g.,][]{Morishita2022}. Finally, although all candidates show prominent Lyman breaks and robust photometric redshift solutions in the expected range, spectroscopic confirmation is needed to rule out any possible contamination in the sample from rare classes of lower-redshift interlopers \citep[][]{Vulcani2017,Zavala2022,Fujimoto2022,Naidu2022c,ArrabalHaro2023}, in particular for objects with limited coverage at short wavelengths.

In conclusion, our study confirms the surprisingly high density of bright
galaxies observed in early {\it JWST} observations, but calls for both extensive follow-up spectroscopy and wider surveys along multiple lines of sight to achieve an unbiased estimate of their number density and clustering properties.

\begin{acknowledgments}
We thank the anonymous referee for their detailed and constructive report. This work is based on observations made with the NASA/ESA/CSA {\it James Webb Space Telescope (JWST)}. The data were obtained from the Mikulski Archive for Space Telescopes at the Space Telescope Science Institute, which is operated by the Association of Universities for Research in Astronomy, Inc., under NASA contract NAS 5-03127 for {\it JWST}. These observations are associated with program JWST-ERS-1324. The {\it JWST} data used in this paper can be found on MAST: http://dx.doi.org/10.17909/fqaq-p393. We acknowledge financial support from NASA through grants JWST-ERS-1342. K.G. and T.N. are grateful for support from Australian Research Council Laureate Fellowship FL180100060. C.M. is supported by the VILLUM FONDEN under grant 37459. The Cosmic Dawn Center (DAWN) is funded by the Danish National Research Foundation under grant DNRF140. We acknowledge support from INAF Mini-grant ``Reionization and Fundamental Cosmology with High-Redshift Galaxies," and grants PRIN-MIUR 2017WSCC32 and 2020SKSTHZ. A.V.F. received financial assistance from the Christopher R. Redlich Fund and numerous individual donors.
\end{acknowledgments}

%% To help institutions obtain information on the effectiveness of their 
%% telescopes the AAS Journals has created a group of keywords for telescope 
%% facilities.
%
%% Following the acknowledgments section, use the following syntax and the
%% \facility{} or \facilities{} macros to list the keywords of facilities used 
%% in the research for the paper.  Each keyword is check against the master 
%% list during copy editing.  Individual instruments can be provided in 
%% parentheses, after the keyword, but they are not verified.

\vspace{5mm}
%\facilities{}

%% Similar to \facility{}, there is the optional \software command to allow 
%% authors a place to specify which programs were used during the creation of 
%% the manuscript. Authors should list each code and include either a
%% citation or url to the code inside ()s when available.

\software{A-PHOT \citep[][]{Merlin2019}, Astropy \citep{Astropy2013}, EAzY \citep{Brammer2008}, Matplotlib \citep{Hunter2007}, SExtractor \citep[v2.8.6][]{Bertin1996,Guo2013}, \textsc{SCAMP} \citep[][]{Bertin2006}, \textsc{SWarp} \citep[][]{Bertin2002}, zphot \citep{Fontana2000}}

%% Appendix material should be preceded with a single \appendix command.
%% There should be a \section command for each appendix. Mark appendix
%% subsections with the same markup you use in the main body of the paper.

%% Each Appendix (indicated with \section) will be lettered A, B, C, etc.
%% The equation counter will reset when it encounters the \appendix
%% command and will number appendix equations (A1), (A2), etc. The
%% Figure and Table counter will not reset.

%\appendix

%\section{Appendix information}

\bibliography{biblio3}{}
\bibliographystyle{aasjournal}

%% This command is needed to show the entire author+affiliation list when
%% the collaboration and author truncation commands are used.  It has to
%% go at the end of the manuscript.
%\allauthors

%% Include this line if you are using the \added, \replaced, \deleted
%% commands to see a summary list of all changes at the end of the article.
%\listofchanges

\end{document}